\documentclass[graybox]{svmult}

\usepackage{type1cm}        
\usepackage{makeidx}         
\usepackage{graphicx}        
\usepackage{multicol}        
\usepackage[bottom]{footmisc}

\usepackage{hyperref}       

\usepackage{newtxtext}       %
\usepackage[varvw]{newtxmath}       
\usepackage{natbib}

\makeindex             

\begin{document}

\title*{CMB Polarization Measurements}
\author{Ettore Carretti and Carlo Baccigalupi}
\institute{Ettore Carretti \at INAF Istituto di Radioastronomia, Via Gobetti 101, 40129 Bologna, Italy, \email{carretti@ira.inaf.it}
\and Carlo Baccigalupi \at SISSA, International School for Advanced Studies, Via Bonomea 265, 34136 Trieste TS, Italy, \email{carlo.baccigalupi@sissa.it}}
%
%
\maketitle

\abstract{
The polarization of the Cosmic Microwave Background (CMB) radiation carries essential information on early stages of the Universe such as the cosmic inflation, forming cosmological structures through gravitational lensing, and the epoch of re-ionization. The signal requires high sensitivity instruments with a large number of detectors (bolometers) and low leakage of Stokes $I$ into $Q$ and $U$.  The Galactic diffuse foreground emission is a limiting factor in CMB polarization measurements, requiring its characterization at both low and high frequency compared to the peak of the CMB emission, in order to be subtracted off.  In this paper we describe the next generation space experiment for the measure of the CMB polarization, LiteBIRD, that is aimed to investigate the first fractions of a second of the Universe and is expected to be flown at the beginning of the next decade. Also, we describe the experiments designed for measuring the foreground emissions from our own Galaxy. Finally, we also describe sub-orbital experiments, operating and planned, as they are vehicles for the development of technologies and data reduction tools that have been and will be used in space missions. }
 
\abstract*{}

\section{Introduction}
\label{sec:intro}

The Cosmic Microwave Background radiation (CMB) is the relic emission of the Big Bang and carries the signature of processes in the very early Universe. The study of its total Intensity  anisotropies, reaching the arcminute angular scales by experiments like the Cosmic Background Explorer (\textit{COBE}), the Wilkinson Microwave Anisotropy Probe (\textit{WMAP}), and \textit{PLANCK} have achieved high precision measurements of cosmological parameters such as the mean density of the Universe, its expansion rate, and the share of baryonic matter, dark matter, and dark energy \citep{2020A&A...641A...6P}. 

Despite the excellent results in polarization measurement by \textit{PLANCK} and \textit{WMAP} in particular, a complete measurement of CMB polarization is the current main goal of CMB observations.  One of the two components of the polarization tensor, the $E$--mode, is essential to reach the current precision on cosmological parameters, and in particular, it is very relevant to investigate the reionization history of the Universe, caused by the UV radiation from the first stars and galaxies at the end of the dark ages \citep{1997PhRvD..55.1830Z}. The other component, the $B$-mode, looks even further back in time. It is dominated by two primary mechanisms: the gravitational lensing produced on background CMB anisotropies from forming cosmological structure, at arcminute angular scales, and the gravitational wave background emitted by the inflation when the Universe was a tiny fraction of a second old, at degree angular scales, also exciting the reionization, occurring at even larger scales \citep{1997PhRvD..55.1830Z}.
 
\begin{figure}
\centering
\includegraphics[width=0.8\hsize]{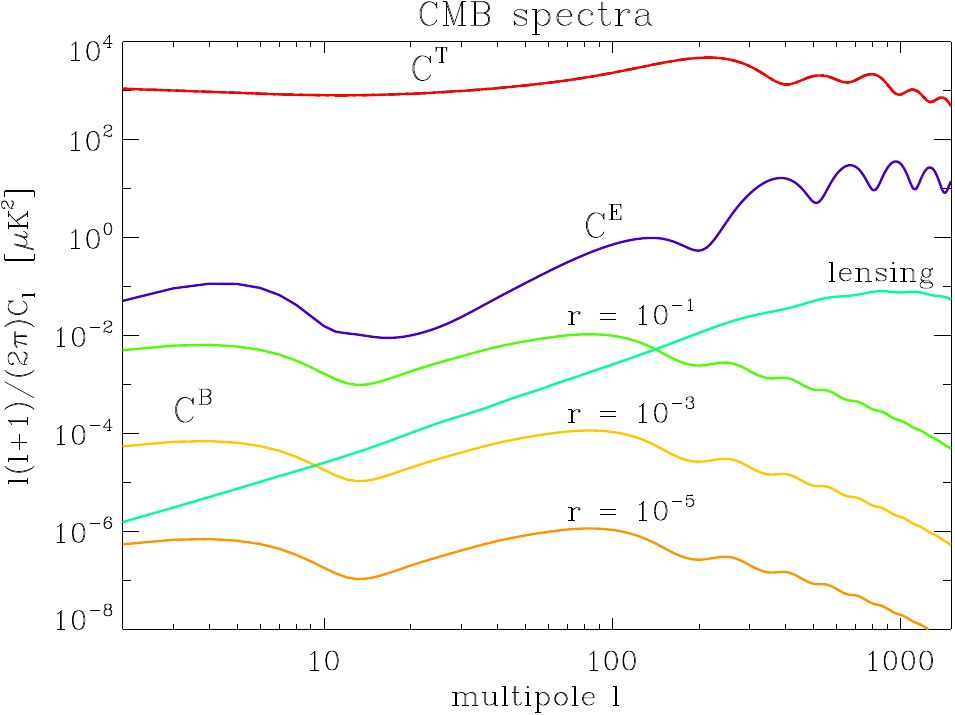}
\caption{CMB angular power spectra of total intensity ($C^T$),  polarization $E$--mode ($C^E$), and $B$--mode ($C^B$) as expected from the current cosmological best fit, the $\Lambda$ Cold Dark Matter model ($\Lambda$CDM). The $B$-mode is shown for three values  of the tensor-to-scalar perturbation power ratio $r$, which measures the power of the primordial inflationary  gravitational wave background. The contribution  to $C^B$ by gravitational lensing is also reported. The CMB has an intrinsic limit of $\Delta r \approx 10^{-5}$, corresponding to a $B$--mode of $\approx$~1~nK. 
}
\label{fig:1}       
\end{figure}

Despite this cutting-edge science, the CMB polarization is weak and a tiny fraction of the total intensity anisotropies (Figure~\ref{fig:1}). The $E$--mode is about 0.3~$\mu$K on large angular scales ($\approx 1$\ \% of the anisotropies).
The most precise measurements that also include the largest angular scales, 90$^\circ$, are from \text{PLANCK} \citep{2020A&A...641A...5P}, while the Background Imaging of Cosmic Extragalactic Polarization/Keck  (BICEP/Keck, \citealt{2022ApJ...927...77A}, see the  \href{http://bicepkeck.org/}{BICEP/Keck web page}), has given the best results on small angular scales \citep{2022arXiv220316556B}. 
The $B$-mode is even weaker and while the lensing power component has been measured in several observations, only upper limits exist for the cosmological gravitational wave component thus far. The latter is  usually measured by  its ratio with  the density perturbations at large scales, the so called tensor-to-scalar perturbation power ratio ($r$). The current upper limit is $r < 0.036$ at 95\% confidence level, from BICEP/Keck \citep{2022arXiv220316556B}.  Signal intensity  values range from the present upper limit of $\approx 0.1$~$\mu$K 
down to the CMB intrinsic limit of $\approx$1~nK, more than four orders of magnitude smaller than the total intensity anisotropies
\citep{2005PhRvD..72l3006A}. Currently, the intensity of the $B$-mode, and hence the corresponding value of $r$, cannot be predicted. The dependence is on the specific model of inflation and on the energy scale  it occurred at. Yet, setting new upper limits is important, because it helps exclude sets of models and set upper limits of the cosmic inflation energy scale. 

The cosmological $B$-mode signal peaks at two different angular scales. The first is the reionization peak at very large angular scales (multipoles $\ell \le 5$, corresponding to scales of $\theta \ge 35^\circ$), the other is at $\theta =2^\circ$, or multipole $\ell =90$. 

As we anticipated, the lensing $B$-mode is generated by the gravitational lensing of the CMB polarization produced by the galaxies of the Large Scale Structure (LSS) of the Universe. While this broad peak, caused by the leakage of the  $E$- into $B$-mode by gravitational lensing,  is a contaminant of the primordial $B$-mode for models with $r$ of the order of 0.01 or less, it is still extremely important to measure other cosmological components, such as the Dark Energy and Matter (DE, DM). To some extend, the lensing can be predicted and subtracted off (de-lensed) using internal CMB lensing reconstruction, or the LSS as measured by other probes \cite{2023arXiv231205194N}

The low signal and low polarization fraction of the CMB polarization require specifically designed experiments: low noise detectors to reach the required sensitivity; high polarization purity to  minimize leakages from the unpolarized component; highly stable environment to minimize  contamination from the surrounding environment  and deliver 
high thermal stability. The latter requires a quiet space environment like the Sun-Earth Lagrangian point 2 (L2), where the telescope can be efficiently shielded from Sun, Earth, 
and Moon ensuring stable conditions. The minimum contamination from diffuse foreground emissions from our own Galaxy occurs at 70--100 GHz \citep{2018A&A...618A.166K},  which sets the best CMB frequency range right where radio and bolometric technologies overlap. Currently, bolomerters have been extended down to 30--40 GHz and are commonly used in most CMB experiments  because of their batter sensitivity. However, the need to trace foregrounds requires observations both at lower and higher frequencies, which requires both these two technologies to cover all of the needed broad frequency range (see Section \ref{sec:foregrounds}). 

This chapter is organized as follows. In Section \ref{sec:suborbital} we discuss sub-orbital  experiments (ground-based or balloon-borne) aimed to develop the required technologies to be ported to the space experiments. The ground projects to measure the  foreground emission at low frequency are discussed in Section \ref{sec:foregrounds}. In Section \ref{sec:prespace} past space experiments are described. Finally, in Section \ref{sec:space} we discuss LiteBIRD, the next generation  space experiment aimed to measure the polarization of the CMB. 

\section{Sub-orbital experiments}
\label{sec:suborbital}

High elevation ground-based and stratospheric balloon-borne experiments are essential to develop technologies that will then be used in space experiments.  Thanks to their large number of detectors they also have their own ambitious scientific goals. Compared to space missions, these experiments suffer from ground and atmospheric contamination that requires mitigation. The cosmic signal is dominated by the Galactic foreground emissions and these experiments are also essential to develop techniques to separate the cosmic signals from each individual source (Component Separation). \textit{Planck},  \textit{WMAP}, and S-PASS data indicate that, over large fractions of the sky, the lowest foreground contamination is in the range 70--100 GHz.  On the other hand, since the signal is foreground dominated, observations in a larger frequency range are needed for characterization and subtraction. The sensor technology is of  two categories: bolometers only, and hybrid, where bolometers at high frequencies are sided by High Electron Mobility Transistor (HEMT) amplifier radiometers at low frequency. 

Radiometric receivers are based on direct amplification at the frequency of the signal.  Usually, Low Noise Amplifiers (LNA) of HEMTs are used because of their high sensitivity as cooled down to cryogenic temperatures of typically  10--20~K.
However, HEMT-LNAs  are affected by gain instabilities that add an extra term to the white noise. This is called 1/$f$ noise after its power spectrum \citep{wollack95}.  Its contribution increases with the integration time $\tau$ instead of decreasing as $\tau^{-1/2}$, preventing any benefit of long exposures. Key parameter is the knee frequency, that is the frequency at which the 1/$f$ noise component equals the white noise and  defines the longest useful integration time ($\tau_{\rm max} \approx 1/f_{\rm knee}$). Total power receivers feature $f_{\rm knee}$ ranging from 100~Hz to 1~kHz resulting in too short integration times.
Correlation receivers correlate signals amplified by different LNAs and largely reduce the common mode and, in turn, the fluctuations induced by gain instabilities \citep{cortiglioni06}.
In this case, $f_{\rm knee}$ is typically of the order of   0.01~Hz, which can keep the instrument stable for hundreds of seconds. That is sufficient because  with an  appropriate scanning strategy (e.g., by spinning the instrument)   the signal is modulated  on time scales ranging from 1~min to 2~min, which enables the data reduction software to remove instability effects on longer time scales.

Bolometers are much more sensitive than  HEMTs, but they require to be cooled down to below 0.3 K, the temperature at which their material turns superconductive. This is obtained with liquid  $^4$He and $^3$He. The 1/$f$ noise of bolometers is more benign than that of HEMTs and the knee frequency is of the order of 0.01 Hz. Special reduction systems are not necessary and the scanning strategy is sufficient, as discussed above. 

There are a number of sub-orbital experiments under development or proposed (e.g., see a summary in \citealt{lspe}). Here we describe a few of them, their technology, and their science goals. They are operating, forthcoming, or scheduled for operations in the next decade. This is not a complete review of all operating facilities, but just those in which one of  the authors has  direct participation, and BICEP/Keck, which is currently driving constraints on $r$. The others, very relevant and important, are represented mainly by the Atacama Cosmology Telescope (ACT) \cite{2023arXiv230405203M}, see the \href{https://act.princeton.edu/}{ACT web  page},
the South Pole Telescope (SPT) \cite{2023PhRvD.108b3510B}, see the \href{https://pole.uchicago.edu/}{SPT web page}, and the proposed European Low Frequency Survey (ELFS) \cite{2023arXiv231016509M}, and South Pole Observatory (SPO). As we already anticipated, the BICEP/Keck array currently gives the best limit on the tensor-to-scalar power ratio, as reported in Section \ref{sec:intro}. 

\subsection{LSPE}

\begin{figure}
\centering
\vskip -0.cm
  \includegraphics[angle=0,width=1.0\hsize]{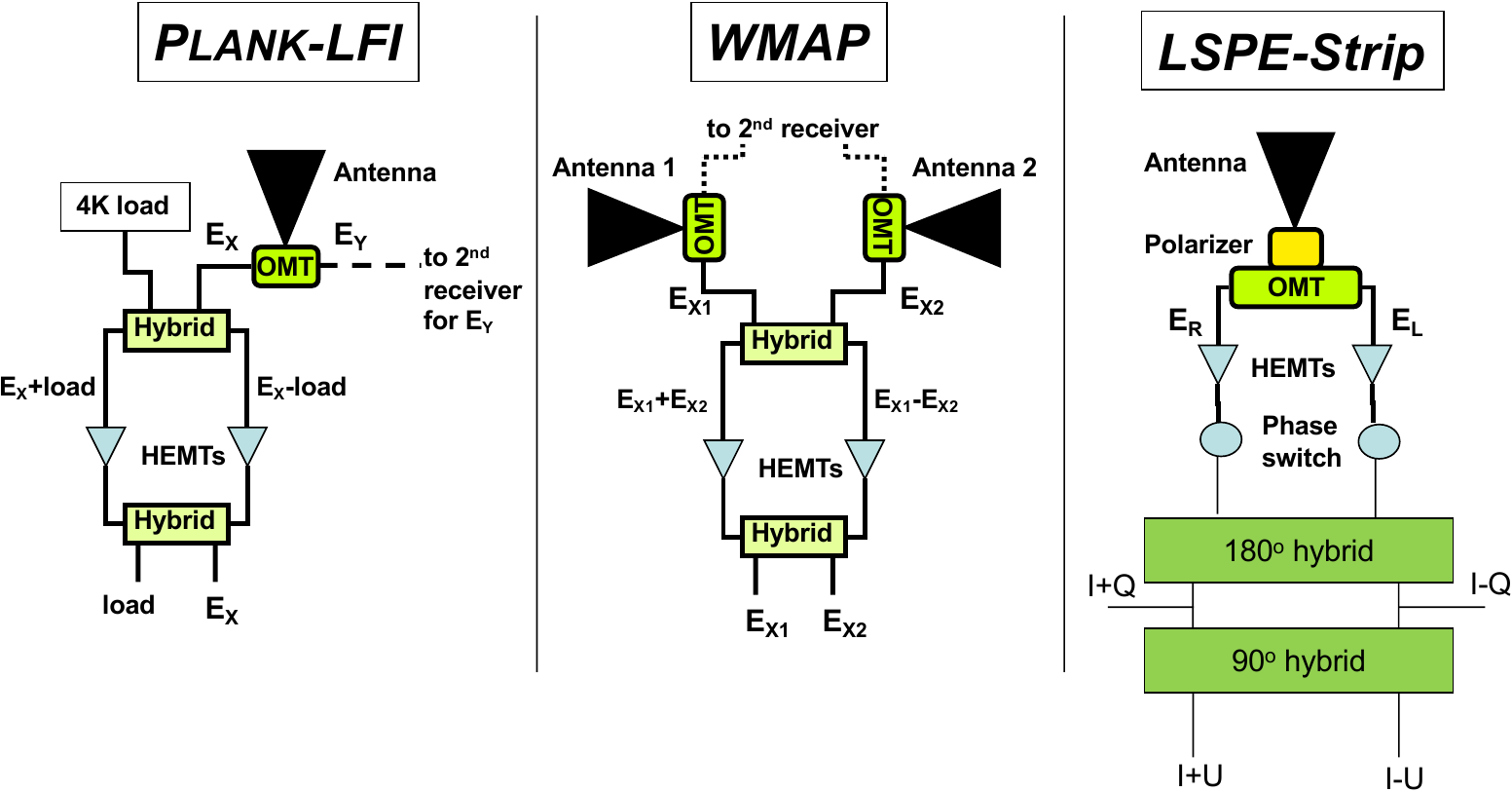}
\vskip -0.cm
\caption{Receiver architectures  of {PLANCK--LFI}, {WMAP}, and {LSPE-Strip}.}
\label{fig:arch}       
\end{figure}

The Large Scale Polarization Explorer (LSPE) \citep{lspe} is a hybrid project expected to observe in the next few years. It consists of two experiments. LSPE-Strip is radiometer-based and its main channel at 43 GHz, in combination with other lower frequency projects, aims at realizing a spatial and frequency behaviour characterization of the Galactic synchrotron emission to then subtract it at the frequencies best suited for a CMB detection.  It will observe from ground at the Teide Observatory, Tenerife, Spain. The scanning strategy is made of azimuthal scans at 1 rpm, which minimise the atmospheric fluctuations. The Earth rotation in one sideral day allows to  cover the entire 24-h R.A. strip. Assuming a zenith angle of $\beta = 35^\circ$, it covers Dec = [$-12^\circ$, $68^\circ$], including the $10^\circ$ receiver array Field Of View (FoV). The telescope is an off-axis, Dragonian configuration with a large FoV, flat focal plane, and low total intensity to polarization leakage. In order to minimise the ground emission contamination the structure is surrounded by an aluminium shield, while the optics has a baffle  with a radiation absorber coating. 
The sensor section is made of 49 receivers. Each of them is a full correlation receiver where both Stokes $Q$ and $U$ are simultaneously detected as correlated, stable outputs. Fig. \ref{fig:arch}, right panel, shows the schematic architecture. A horn feed antenna collects the radiation from the optics and a polarizer and an OMT (OrthoMode Transducer) extracts the two  Left- and Right-Hand  circular polarizations. After the amplification by HEMTs, a 180$^\circ$ hybrid gives two outputs that, once detected, are $I \pm Q$ that, once subtracted, give $Q$. A further 90$^\circ$ hybrid gives  $I \pm U$ as outputs  that, once subtracted, give $U$. 
Phase switches between amplifiers and hybrid switch the signs of $Q$ and $U$, which, once demodulated by the instrument electronics, remove the leakage of Stokes I into the polarized signal produced by the hybrid. The leakage of Stokes $I$ into $Q$ and $U$ is a contaminant term of CMB instruments. This correlation receiver has excellent polarization purity with a leakage better than -20 dB (1 percent), mostly by the OMT. All the receiver is cooled down to a temperature of 20 K in order to minimise the noise generated by HEMTs and passive components.  Assuming a 2-yr observation, the map sensitivity is 130 $\mu$K arcmin. The angular resolution is 21 arcmin. 

LSPE-SWIPE (Short-Wavelength Instrument for the Polarization Explorer) is a bolometer-based, balloon-borne instrument to be flown in a winter arctic flight at some 35 km of altitude. It consists of three channels at 145, 210, and 240 GHz. The first channel is for CMB measurements, while the other aim at the characterization of the Galactic dust emission. The launch can occur from either Svalbard or Sweden for an up to 15-day flight. A winter arctic flight is all by night, which allows a full spin of the telescope and thus the observation of a large 24-h R.A. strip, as for the other instrument. The residual atmosphere at 35-km of altitude is little, which makes the contamination negligible. The zenith angle is changed in flight in the range $\beta$ = [35$^\circ$, 55$^\circ$]  so that the observed area covers Dec = [$13^\circ$, $77^\circ$], assuming a flight from Svalbard and including the $20^\circ$ FoV.  The common area with Strip is about 35\% of the sky.  The optics has a baffle with a radiation absorber coating. 
The sensor section is made of 162, 82, and 82 Transition Edge Sensor (TES) bolometers at the three frequencies. The instrument is made of a lens that converges the radiation to the bolometers. Between lens and bolometers, a grid polarizer tilted  by 45$^\circ$ splits the two linear  polarizations $E_X$ and $E_Y$ in two directions. The transmitted one passes through and goes to a focal plane with a half of the bolometers, while the reflected radiation is bent by 90$^\circ$ and hits a second focal plane with the other half of the bolometers.  The difference of the power detected by the two sets of bolometers gives Stokes $Q$, albeit with a poor leakage rejection. The bolometers are fed with a feed horn each to improve their coupling with the radiation. The feeds are multi-mode to increase the number of  collected propagation modes and, hence, the sensitivity. This comes at the expense of the angular resolution, but that it is not an issue at the required angular scales. To obtain Stokes $U$ and improve the leakage rejection, a spinning half wave plate (HWP) is placed skywards the lens. The HWP produces birefringence and modulates  $Q$ and $U$ of the  sky polarized radiation  as a function of  its spin angle $\varphi$ as 
\begin{eqnarray}
    Q^\prime &=& +Q \cos(4\varphi) + U \sin (4\varphi) \\
    U^\prime &=& -Q \sin(4\varphi) + U \cos (4\varphi).  
\end{eqnarray} 
The leakage term, proportional to Stokes $I$, is not modulated and can be efficiently rejected. The residual leakage is $< -20$ dB thanks to the HWP.   The instrument, including the HWP, is cooled down to a temperature of 1.6 K to minimise the thermal noise.  The bolometers are cooled down to 0.3 K, at which the material they are made from is superconductive.  Assuming a 15-day observation, the expected map sensitivity is 10 $\mu$K arcmin, which explains why the bolometers are the preferred sensors at viable frequencies. The angular resolution is 85 arcmin.  

Thanks to the large covered area and the resolution, LSPE can access both peaks of the cosmological $B$-mode. The final sensitivity on $r$ of the two combined instruments, including the errors of the foregrounds subtraction, is of $\sigma_r < 0.01$, or an upper limit of $r < 0.015$ at 95\% confidence level. 

\subsection{BICEP/Keck Array}

BICEP/Keck Array \citep{2022arXiv220316556B} has observed for a decade and will observe until the end of the present decade. It is a series of ground-based telescopes located at the Amudsen-Scott South Pole Station  and that have observed at different times. The location gives 6-month nights at low temperature (-70 C), which ensures a stable atmosphere with low content of water vapour, the major emitter at mm-wavelengths. The telescopes are protected by a reflective shield and have a baffle with an absorber coating. BICEP2, the first to observe at the beginning of 2010s, is a TES bolometer-based system with a single channel at 150 GHz. Its scheme is the base for all the other telescopes. It is a simple architecture. A 250-mm diameter lens converges the radiation to 500 bolometers that are cooled at 0.25 K. Each bolometer of a pair has an antenna sensitive to one polarization, orthogonal for the two pair sensors. Their difference gives Stokes $Q$. To get both Stokes $Q$ and $U$ the entire telescope can be rotated in steps and the observing time is divided in runs at a fixed angle each. The Keck Array was initially an array of five BICEP2s to improve the sensitivity, then converted to two channels at 95 and 220 GHz to map and subtract the Galactic dust foreground emission, that, along with the Galactic Synchrotron emission, dominates the signal at the CMB frequencies. BICEP3 has observed from 2016 and replaced BICEP2. It is a larger version of the latter with a 500 mm diameter lens to have a larger focal plane that accommodates 2500 bolometers at 95 GHz, in the range of minimum foreground emission. These three telescopes have observed a field of size of some 2\% of the sky, aiming to measure the 2$^\circ$ peak of the $B$-mode. The observations conducted till 2018 were used to obtain the currently deepest observation ($\sigma_r < 0.009$) with an  upper limit of $r < 0.036$ at 95\% confidence level \citep{2022arXiv220316556B}. 

The Keck  Array has been decommissioned in 2019 and a replacement with  the BICEP Array is expected to observe till the end of the decade. The BICEP  Array is made up of four BICEP3-like receivers, except it accommodates six frequency channels at 30, 40, 95, 150, 220, and 270 GHz to also measure the diffuse Galactic synchrotron and dust emission. The number of bolometers is  30,000, an order of magnitude more than BICEP3. It is expected to reach a sensitivity on $r$ of $\sigma_r < 0.003$.

\subsection{PolarBear/Simons Array}

A complete description of the PolarBear/Simons Array (PBSA) project can be found in the recent paper by the collaboration \cite{2023PhRvD.108d3017A} and the \href{http://bolo.berkeley.edu/polarbear/}{PBSA web page}. 

The instrument is made of a system of three 2.5 m Huan Tran reflective Telescopes (HTT) containing cryogenic detectors, located at the James Ax Observatory in the Atacama Desert in Chile
at an elevation of 5,190 m. The optics is equipped with modulation (by a HWP) of the polarization on the sky with respect to the receiver. The receivers contain approximately $10^{3}$ Transition Edge Sensitive (TES) bolometers on 7 wavers on the focal plane, which was cooled to 0.3 K, set for observations at 95, 150, 220 GHz with 3.5 arcmin angular resolution. Detectors are organized in wafers, interfaced with readout electronics constituted by pixels which, in the PBSA focal plane architecture, are sensitive to both $Q$ and $U$ Stokes parameters. 

The collaboration consists in hundreds of collaborators and laboratories in the Unites States, Japan, United Kingdom and Europe. The project is supported by the National Science Foundation, the Simons Foundation, the Templeton Foundation, the Department of Energy at the Lawrence Berkeley National Laboratory, as well as the James Ax Observatory facilities in Atacama, and grants and projects from Japanese, European (both Union and National), United Kingdom and Australian agencies. 

Science operations have been accomplished so far through one PBSA Telescope, the PolarBear, achieving the first detection of B-modes based on auto power spectra, several measurements of lensing spectrum, reconstruction and de-lensing, as well as an upper limit on the cosmological $r$ based on degree scale B-mode measurements, as well as several other science results, see  \cite{2023PhRvD.108d3017A} and references therein. 

\subsection{Simons Observatory}

A complete description of the Simons Observator (SO) project can be found in the recent paper by the collaboration \cite{2019JCAP...02..056A} and the \href{https://simonsobservatory.org/}{SO web page}. 

SO consists of a Large (6 meters) Aperture Telescope and a system of three Small (0.5 meters) Aperture Telescope (LAT, SATs), located in the Atacama desert, starting science operations in 2024. A number of  3 more LATs will be constructed by 2028, one from Japan and two from the United Kingdom. The LAT will be observing about 40\% of the sky in six frequency bands (27, 39, 93, 145, 225, and 280 GHz) with  angular resolution ranging from 7 to 1 arcmin. The LAT receiver will have 30,000 TES bolometric detectors distributed among seven optics tubes spanning the six frequency channels. Each LAT
tube will contain three arrays, each on a detector wafer, each measuring two frequency bands and in  two linear polarizations. The SATs will measure 10\% of the sky with degree angular resolution. The receivers in the SATs are interfaced with a single optics tube for each telescope, and will together also contain 30,000 detectors. The SAT optics
tubes will each house seven arrays, and will each have 
a continuously rotating HWP. 


The collaboration consists in nearly 400 collaborators and several laboratories in the Unites States, Japan, United Kingdom and Europe. The project is primarily supported by the Simons Foundation, with additional support from the Heising-Simons Foundation, the National Science Foundation in the United States (NSF), national funding agencies in the United Kingdom, Europe,
and Japan, and member institutions. 

Four telescopes are starting science operations in 2024. The primary goal is constituted by the measure of the tensor-to-scalar power ratio from degree scale $B$-modes with $3\times 10^{-3}$  1$\sigma$ error. The accuracy of polarization maps, at least one order of magnitude better than PLANCK, will allow to probe with unprecedented precision the Gaussianity of primordial perturbations, the Big Bang Nucleosynthesis through primordial Helium, cosmological gravitational lensing, the nature and interactions of dark energy and matter in cross-correlation with optical surveys, constraining the ionization efficiency in models of reionization, detecting tens of thousands of galaxy clusters and extra-Galactic objects, studying the diffuse gas in our own Galaxy. 

\subsection{CMB-S4}

A complete description of the CMB Stage 4! (CMB-S4) project can be found in the recent paper by the collaboration \cite{2022ApJ...926...54A} and the \href{https://cmb-s4.org/}{S4 web page}. 

S4 will be a network of detectors operating in Atacama and South Pole. A deep survey, targeting measurement of $r$ through $B$-mode measurements will be achieved using fourteen 0.55 meter refractor SATs (at 155 GHz and below) and four 0.44 meter SATs (at 220/270 GHz), with dichroic, TES detectors, in each SAT measuring two of the eight targeted frequency bands between 30 and 270 GHz, as well as one 6 meters class crossed-Dragone LAT, equipped with detectors distributed over seven bands from 20 to 270 GHz. A deep and wide survey covering approximately 70\% of the sky to be conducted using two 6 meters crossed-Dragone LATs located in Chile. 
The total detector count for the eighteen SATs surpasses $1.5\times 10^{5}$ with the majority of the detectors allocated
to the 85 to 155 GHz bands. The SATs will have 204 wafers. 
LAT will have a total TES detector count over  $1.1\times 10^{5}$, with the majority of the detectors allocated to the 95 to 150 GHz bands, organized in 76 wafers. At the South Pole, the polarization signal is measured by differencing pairs of orthogonal detectors, modulated by scanning the telescope in azimuth. Any telescopes deployed to Chile at 85 GHz and above will be equipped with HWPs. The atmosphere
is sufficiently stable in the lowest frequency bands that a HWP is not needed. In the reference design, the total detector count for S4 is 511,184, and it requires 432 science grade wafers. 

The collaboration consists in hundreds of collaborators and laboratories in the Unites States, Japan, United Kingdom and Europe.
CMB-S4 will be a joint agency program with comparable support from the Department of Energy and the National Science Foundation, receiving strong recommendation lately in the Report of the \href{https://usparticlephysics.org/2023-p5-report/}{2023 Particle Physics Project Prioritization Panel}. 

The instrument is scheduled to start observations early in the next decade. The first goal is to measure $r$  larger than $3\times 10^{3}$ with 5$\sigma$ accuracy. 
The second goal is to detect or strongly constrain departures from the thermal history of the Universe predicted by the Standard Model of particle physics, through the effective number of
relativistic species in the early Universe. S4 will constraint a departure from the effective number corresponding to 6\% at 1$\sigma$ confidence level. The third goal is represented by the corresponding survey of a large fraction of the sky at centimeter and millimeter wavelengths, measuring gravitational lensing and investigating the nature of dark energy and matter and their potential interactions with known particles of Standard Model of Particle Physics, neutrino masses, revealing tens of thousands of galaxy clusters and extra-Galactic objects, in cross-correlation with surveys in other bands, as well as the properties of the diffuse gas in our own Galaxy \cite{2022ApJ...926...54A}. 

\section{Foreground dedicated experiments}\label{sec:foregrounds}
The synchrotron and dust emission are  usually measured at low and high frequency with respect to the minimum of foreground emissions, respectively. The smaller number of sensors imposed  by the larger size of  low-frequency receivers makes the experiments less sensitive at these frequencies. Also, the frequency behaviour has large spatial variations, requiring a large frequency range in order to characterise and obtain a precise extrapolation of synchrotron to the CMB frequencies. The Galactic synchrotron emission measurements are conducted from the ground, using radio telescopes equipped with HEMT-based  receivers at 2--20 GHz. The steep brightness temperature spectrum  makes their signal-to-noise ratio favourable. Here we present the all sky class projects that can provide CMB probes  with accurate synchrotron measurements:  S-PASS at 2.3 GHz, C-BASS at 5 GHz, and QUIJOTE at 10--40 GHz. 

\subsection{S-PASS}
S-PASS (S-band Polarization All Sky Survey) \citep{2019MNRAS.489.2330C} is a project that has mapped the linear polarization of the entire southern sky up to Dec = -1$^\circ$ at 2.3 GHz and with an angular resolution of 8.9 arcmin. It was observed in 2007--2010 using Murriyang, the 64-m radio telescope at Parkes, NSW, Australia.  The frequency was chosen because high enough to limit Faraday rotation (FR) effects (it depolarizes  the emission at Galactic latitudes $|b| < 30^\circ$ at 1.4 GHz) while low enough to maximise the signal-to-noise ratio $S/N$. The results actually showed that the mid- and  high-latitude sky has little FR effects, while 98\% of the observed sky has $S/N >3$ ($S/N >20$ at a resolution of $1^\circ$), which makes S-PASS the Galactic synchrotron foreground survey with the best $S/N$ currently available or planned. The observing strategy is based on long azimuth scans at the elevation of the south Celestial pole at Parkes, which minimises ground emission and atmospheric variations. The project used the Galileo receiver. A feed horn collects the radiation from the parabolic primary mirror, and a polarizer and an OMT extracts the two circular polarizations. A digital correlator performs  their cross-product, whose Real and Imaginary parts are $Q$ and $U$, respectively: 
\begin{equation}
    Q+iU = 2\,E_RE_L^*\, . \label{eq:qucross}
\end{equation}
The residual leakage after calibration is better than 0.2\%.  The foreground analysis  done with this dataset in combination with PLANCK's has shown that the polarized signal intensity and spectral behaviour have large spatial variations \citep{2018A&A...618A.166K}. This  means that mean intensity and spectral behaviour are insufficient and calls  for a precise mapping of this foreground to subtract it and obtain high-precision $B$-mode measurements. Many more astrophysical studies have been conducted with this dataset, from Galactic astrophysics to galaxy clusters and  the cosmic web (see \citealt{2019MNRAS.489.2330C} for a summary). 

\subsection{C-BASS}
The C-BASS (C-Band All Sky Survey) \citep{2018MNRAS.480.3224J} has mapped the entire sky at 5 GHz, at a resolution of 45 arcmin, in total intensity and polarization. Here we described the status of the polarization only. The project is divided in two experiments, each with their telescope, one in the north at the Owens Valley Radio Observatory, California, and the other in the south at Klerefontein, South Africa. The observations have been completed for the northern survey. The FR effects are smaller at 5 GHz compared to S-PASS. However the signal and the $S/N$ are  weaker, and C-BASS and S-PASS are somewhat complementary. The weaker signal makes the ground emission contamination more relevant. In order to reduce it, the northern telescope is equipped with absorbing baffles at the mirror edges. The southern telescope is larger and the feed horn under illuminates the primary mirror, so that the ground at the mirror edge is picked at lower levels and baffles are not necessary. The observing strategy is based on azimuth scans to minimise the ground emission  variations, as for S-PASS. 

The receivers of the two experiments are identical down to the backends. A feed antenna gets the radiation from the optics. A waveguide OMT extracts the two linear polarizations that are then converted to the two circular polarizations by a Linear-to-Circular (L2C) converter planar circuit. For the sake of the stability of the total intensity output, thermal loads ($L_1$ and $L_2$) are added to the system after the L2C by two 180$^\circ$ hybrids (one per polarization state), whose four outputs are $E_R \pm L_1$ and $E_L \pm L_2$. The four signals are then amplified by HEMTs. The circuit is quite complex and well described by Fig. 4 of \citet{2018MNRAS.480.3224J}. The two experiments are then designed differently. The southern station uses a digital correlator and $Q$ and $U$ are obtained as cross-products of $E_R$ and $E_L$ (Eq. \ref{eq:qucross}). The northern station does the same product, except for the fact that it uses an analogue correlator based on a  90$^\circ$ hybrid. The total intensity outputs are obtained by separating again and subtracting the two polarizations and the two loads. Sky signals and loads go through the same amplification chains, which makes their differences much more stable with respect to the standard total power outputs of a radio receiver. As in LSPE-Strip, phase switches are added to similar goals. The feed is partly cryogenically cooled. The rest of the receiver up to just before the backend is cooled down to 4 K (north) and 10 K (south).

\subsection{QUIJOTE}

QUIJOTE (Q-U-I JOint Tenerife Experiment)  is a ground-based experiment at the Teide Observatory, Tenerife, Spain. It has been and observing the Galactic synchrotron foreground of the northern sky down to Dec $\approx -30^\circ$ at 10--20 GHz   \citep{2023MNRAS.519.3383R, 2016SPIE.9906E..1KP}. QUIJOTE consists of two telescopes and three instruments. The telescope optics is an offset Dragone configuration with no obstruction and small leakage ($<-25$ dB). The instruments are the Mid Frequency Instrument (MFI) at 10--20 GHz, the Thirty GHz Instrument (TGI), and the Forty GHz Instrument (FGI). The observing strategy is based on azimuth scans by  fast spinning  the telescopes. 

TGI and FGI focus on measuring the CMB $B$-mode with an upper limit of $r=0.05$. MFI is  dedicated to the measurement of the Galactic foreground in four 2-GHz broad bands centred at 11, 13, 17, and 19 GHz. MFI  is made up of four receivers, two at 10--14 GHz giving the two bands at 11 and 13 GHz each, and  two at 16--20 GHz  giving the two bands at 17 and 19 GHz each. The architecture scheme of each receiver (HEMT-based) is well described in Fig. 3 of \citet{2016SPIE.9906E..1KP}. The radiation from  the mirrors is collected by a feed horn and the two linear polarizations $E_X$ and $E_Y$ are split by on OMT. To equally amplify  the two polarizations  a 90$^\circ$ hybrid is placed after the OMT and gives the two outputs $E_X \pm iE_Y$, that are then amplified by HEMT amplifiers. From the OMT to this stage, the system is cooled down to 20 K. The two signals are then further amplified and a second  90$^\circ$ hybrid   gives  $E_X$ and $E_Y$  back. A 180$^\circ$ hybrid polarimeted gives the two outputs $E_X \pm E_Y$ back. Once measured, and subtracted, they give $E_X E_Y^*$, for which the real part is $U$. A splitter before the 180$^\circ$ hybrid polarimeter gives $E_X$ and $E_Y$ that, once detected and subtracted, give $Q$.  he two polarizations have common amplification, thus $Q$ and $U$ are stable with small leakage. 

The analysis of these maps has confirmed the spatial variability of intensity and  frequency behaviour of the synchrotron foreground found with the S-PASS data, and the need of  a precise characterization for efficient subtraction in operating and future B-Mode observations. 

\section{Completed Space based CMB observations}
\label{sec:prespace}

Polarization sensitive satellites, \textit{WMAP} \textit{PLANCK} have been orbiting at the Lagragian point L2 of the Earth-Sun system to benefit of its exceptionally stable  thermal conditions and easy-to-shield solar, Earth, and Moon emissions. WMAP flew in the first decade of 2000s, whilst PLANCK in the 2010s. PLANCK scanning strategy was spinning the spacecraft, whilst WMAP added a precession to add more crossing between scans. PLANCK frequency channels spanned the range  30-857 GHz using both HEMTs and bolometers at low and high frequency, respectively. WMAP frequencies spanned 23--94 GHz only using HEMT radiometric receivers. Both experiments  detected  the $E$-mode of the CMB that allowed a first estimate of the optical depth of the epoch of reionization. The instrument and main results of PLANCK are described in \citet{2020A&A...641A...1P, 2020A&A...641A...2P, 2020A&A...641A...3P}, while those of WMAP are in \citet{2003ApJS..148...29J} and \citet{2013ApJS..208...20B}. Here we describe the receiver architectures of these two experiments. 
 
\subsection{\textit{PLANCK-LFI} and \textit{PLANCK-HFI}}

 ESA's \textit{PLANCK} Low Frequency Instrument (\textit{PLANCK-LFI}) consisted of radiometers to cover three frequency bands in the range  30--70~GHz. \textit{PLANCK} was primarily dedicated  to CMB total intensity anisotropies and measured the total power of the two linear polarizations $|E_X|^2$,  $|E_Y|^2$. The sensitivity to both $|E_X|^2$ and  $|E_Y|^2$ implies polarization sensitivity. $Q$ is measured by difference 
\begin{equation}
   Q = |E_X|^2 - |E_Y|^2.
   \label{diffEq}
\end{equation}
Stokes $U$ is obtained by a second receiver, rotated by 45$^\circ$.  The total power is obtained by a pseudo-correlation that  ensures the required  stability. The latter is based on the difference between the sky signal ($\approx$ 2.725~K) and a reference load of comparable intensity: a black body at 4~K, stable thanks to a liquid $^4$He bath.
Each linear polarization collected  by the telescope  is extracted by a feed and an  OMT and combined with the reference load signal by a 180$^\circ$ hybrid (Figure~\ref{fig:arch}, left panel).  
The two outputs, $E^{\rm sky}_X + E_X^{\rm load}$ and $E^{\rm sky}_X - E_X^{\rm load}$ for $E_X$ polarization, are then amplified and combined back  to $E^{\rm sky}_X$ and $E_X^{\rm load}$ by a second hybrid polarimeter. After the detection, sky and load are subtracted off. The two signals are both amplified by the two LNA chains, and their difference cancels out most of the gain fluctuations~\citep{mennella04}. 

The PLANCK High Frequency Instrument (PLANCK-HFI) consisted of  6 bands at 100--857 GHz and was based on bolometers.  It carried 32 PSBs (Polarization Sensitive Bolometers) at four frequency channels, from 100 GHz to 353 GHz. PSBs incorporate  devices to make them sensitive to one linear polarization state \citep{jones02}. At least three  of such detectors are required to measure both $Q$ and $U$. 

\subsection{\textit{WMAP}}

\textit{WMAP} has been observing five bands at 23~GHz to 94~GHz. Besides providing a  measurement of the CMB anisotropy spectrum, it  performed the first detection of the large scale $E$-mode polarization~ term. As \textit{PLANCK}, its primary goal has been the total intensity anisotropy, also being able to measure the polarization by difference of $|E_X|^2$ and $|E_Y|^2$. \textit{WMAP} used differential receivers that subtract  the  signals collected by two independent telescopes looking at two sky positions separated by  140$^\circ$. 
The receiver scheme is similar to that of {PLANCK-LFI}, except  that the signal from the second telescope is subtracted instead of the reference load (Figure~\ref{fig:arch}, mid panel). The other polarization extracted by the two OMTs feeds a second receiver.  The equations to derive Stokes $Q$ and $U$ are given in~\citet{page07}. 

\section{Next Generation space experiment: \textit{LiteBIRD}}
\label{sec:space}
    
Observations from space are demanding in terms of resources, but in particular for cosmological measurements, they have the advantage of accessing the entire sky. In addition, space environments as that of  the Lagragian point L2 of the Sun-Earth system can benefit in terms of reduction of instrumental systematics, being unaffected by atmospheric emission, achieving thermal stability, and no ground pickup, together with enhanced sensitivity due to optimal cryogenic conditions. Here we describe the next generation space mission, \textit{LiteBIRD}, scheduled  for the beginning of the next decade to observe from L2. A complete description of the \textit{LiteBIRD} project can be found in the recent paper by the collaboration \cite{2023PTEP.2023d2F01L} and the \href{https://litebird.isas.jaxa.jp/static/eng/}{LiteBIRD web page}. 

\subsection{Instrument}

\textit{LiteBIRD} (Lite satellite for the studies of B-mode polarization and Inflation from cosmic background Radiation Detection) will be launched on an H3 Japan new rocket, developed by its prime contractor, Mitsubishi Heavy Industries. The first stage of the H3 rocket will adopt the newly developed liquid fuel  engine, more powerful with respect to predecessors in terms of trust. The launch capability of the H3 rocket to a geostationary transfer orbit will be the highest among launch vehicles from the Japanese Space Agency (JAXA), exceeding that of the existing ones. The launch facility at Tanegashima Space Center will also be upgraded following the development of H3. The planned H3 configuration to be used consists in two first stage engines,
two boosters, and a long fairing. The estimated launch capability with this configuration is larger than 3.5 t, resulting in the corresponding requirement on the total weight of the satellite. 

The \textit{LiteBIRD} PayLoad Module (PLM) consists of three telescopes at low, medium, and high frequencies, cooled down to 0.1 K. The PLM also includes the global cooling chain from 300 K to 4.8 K, drivers and warm readout electronics of the detectors. Following a study focusing on the requirement performance, \textit{LiteBIRD} will use as the first optical element a continuously rotating HWP, allowing measuring  polarization from a single detector, suppressing the 1/f noise, and thus distinguishing between the instrumental polarized signal and the sky signal. Without HWP, data from a pair of detectors mutually orthogonal in their polarization orientations are usually combined, causing leakage from temperature to polarization if there are any differences in the beams, gains, or bandpasses between the two detectors. In order to guarantee appropriate thermal performance in terms of stability and minimal heat load, the three telescopes will be equipped with polarization modulator units (PMUs) continuously rotating at a few Hz around a stable temperature below 20 K, using a magnetic levitating mechanism. The distribution and the number of bands over a wide range of frequencies, from 34 GHz to 448 GHz, have been optimized to deal with the control of the diffuse foreground and Carbon Monoxide emissions, the frequency range of HWP materials, control of instrumental systematics thanks to overlapping bands. Therefore, 
a reflective Low Frequency Telescope (LFT) (34–161 GHz), and two refractive Middle and High Frequeency Telescopes (MHFT), the MFT (89–225 GHz) and HFT (166–448 GHz). The MFT and HFT telescopes are mounted on the same mechanical structure, and point to the opposite direction compared to the LFT, but cover the
same circle over the sky when spinning. The focal planes are populated with multi-chroic polarized TES detectors, using lenslet for the LFT and MFT, and horncoupled detectors for the HFT, for a total of 4.5$10^{3}$ units cooled down at 0.1 K. The readout electronics exploits a  frequency multiplexing setup in order to accommodate this large number of detectors without loss of information. The three \textit{LiteBIRD} telescopes are fully cooled down to 4.8 K. All telescopes have intermediate cold stages at 1.75 K and 0.35 K between their mechanical structure at 4.8 K and the detectors at 0.1 K. 

The LFT aperture diameter is of 400 mm and the angular resolution ranges from 24 to 71 arcmin. It consists of nine broad frequency bands spanning 34 to 161 GHz, in order to cover the spectral domains of both CMB and diffuse Galactic emission, dominated at these frequencies by the synchrotron radiation emitted by cosmic ray electrons spiraling around the Galactic magnetic field. 
The LFT optical design follows a crossed-Dragone configuration. The first optical component is  the HWP and the  focal plane is made by TES detectors cooled down to 0.1 K. 

The frequency bands of the MFT range  89 to 224 GHz, while those of HFT  from 166 to 448 GHz. Each telescope features its own polarization modulator, in order to mitigate the bandwidth limitations of the HWP. MFT and HFT assemblies are held on a single mechanical structure cooled down to 4.8 K, corresponding to the MHFT. 

\subsection{Collaboration}

The collaboration includes several laboratories and hundreds of scientists in Japan, United States of America, Europe, United Kingdom. JAXA supports the pre-Phase A2 studies, through the acceleration program of the JAXA Research and Development Directorate, by the World Premier International Research Center Initiative (WPI) of MEXT, by the JSPS Core-to-Core Program, and by JSPS KAKENHI. The Italian \textit{LiteBIRD} Phase-A contribution is supported by the Italian Space Agency, the National Institute for Nuclear Physics (INFN), the National Institute for Astrophysics (INAF), and the Italian Ministry of Foreign Affairs and International Cooperation. The French \textit{LiteBIRD} Phase-A contribution is supported by the Centre National d’Etudes Spatiale (CNES), by the Centre National de la Recherche Scientifique (CNRS), and by the Commissariat à l’Energie Atomique (CEA). The Canadian contribution is supported by the Canadian Space Agency. The US contribution is supported by NASA. Norwegian participation in \textit{LiteBIRD} is supported by the Research Council of Norway. The Spanish \textit{LiteBIRD}
Phase-A contribution is supported by the Spanish Agencia Estatal de Investigación (AEI). Funds that support the Swedish contributions come from the Swedish National Space Agency and the Swedish Research Council. The German participation in \textit{LiteBIRD} is supported in part by the Excellence Cluster ORIGINS, which is funded by the Deutsche Forschungsgemeinschaft (DFG, German Research Foundation) under Germany’s Excellence Strategy. \textit{LiteBIRD} work has received funding from the European Research Council
(ERC) under the Horizon 2020 Research and Innovation Programme. This research used resources of the Central Computing System owned and operated by the Computing Research Center at KEK, as well as resources of the National Energy Research Scientific Computing Center, a DOE Office of Science User Facility supported by the Office of Science of the DOE in the United States of America. 

\subsection{Status}

Observing CMB anisotropies from space relies on the previous experiences, represented by the COsmic Background Explorer (COBE), the Wilkinson Microwave Anisotropy Probe (WMAP) and PLANCK. 
All frequencies are accessible, unlike on the ground where water and oxygen lines block access and reduce the ability to build a detailed model of the foreground emission. Detector sensitivity is higher in space than on the ground due to the absence of atmospheric
loading and the disparity increases rapidly with frequency. The absence of atmospheric emission and its large brightness fluctuations in space-based  measurements gives high fidelity maps on large angular scales. Bright sources such as  Earth and Sun are kept far from the boresight of the telescope by a large angle, giving very low systematic errors due to pickup of those sources in the telescope sidelobes. 

The primary goal of \textit{LiteBIRD} is the measurement of the tensor-to-scalar power  ratio $r$ with a total uncertainty of   $10^{-3}$ or less. This value shall include contributions from instrumental statistical noise fluctuations, instrumental systematics, residual foregrounds, gravitational lensing, and observer bias, and shall not rely on future external data sets. In addition, \textit{LiteBIRD} will achieve full sky CMB linear polarization anisotropy maps for detection of the contribution from tensors on the spectral regions where they are most significant, from the sky quadrupole down to the degree scale, reaching a detection level of   $5\sigma$ assuming an $r$ of $10^{-2}$ and the current values of measured cosmological parameters. 

In addition to the primary goal, the full sky maps in 15 microwave bands will offer rich new data sets, which will enable breakthroughs in a variety of science areas, including the reionization of the Universe, cosmic birefringence, the hot gas in the Universe probed using the Sunyaev–Zel'dovich effect, spatially varying deviations from a perfect Planckian blackbody CMB spectrum, primordial magnetic fields, tests using polarization of the homogeneity in the distribution of total intensity anisotropies, and Galactic astrophysics. 

Finally, the \textit{LiteBIRD} space mission has strong synergy with ground based CMB polarization anisotropy experiments including SO, and S4. Given the small amplitude of $r$ and the disturbances represented by the effects of foregrounds, lensing, and instrumental systematic uncertainties, having highly sensitive data from both LiteBIRD and ground based experiments will contribute to building confidence in the robustness of an inflationary signal detection. Furthermore, \textit{LiteBIRD} provides data on foregrounds at frequencies above the highest frequency that will be observed by the above-mentioned ground-based experiments (280 GHz), and the \textit{LiteBIRD} high frequency data can be combined with those from ground based facilities, improving foreground separation. As well as \textit{LiteBIRD} augmenting ground based experiments, CMB lensing data from the ground can be combined with LiteBIRD data to improve the sensitivity of \textit{LiteBIRD} to $r$ on degree angular scales. Ground based experiments generally focus on a relatively small region of sky, which has the benefit of giving high signal to noise per spatial mode but on a relatively low number of spatial modes. \textit{LiteBIRD}, in contrast, will measure the entire sky and measure all the spatial modes that are
available at moderate signal to noise. Both deep and wide observations can contribute to our understanding of foreground emission. The \textit{LiteBIRD} measurement of the entire sky will probe the variability of the emission in different directions and test the fidelity of the model. There is a similar complementary in terms of instrumental systematic uncertainties, where the high signal to noise ratio per mode of ground based measurements provides a deep probe for discovering weak systematic errors, whereas \textit{LiteBIRD} has the statistical power to measure many spatial modes, in addition to the advantage of making the observations in the benign and stable environment of space. The system represented by ground based efforts and  space observations by \textit{LiteBIRD} will be ideal for mapping out the CMB polarization anisotropies in a definitive manner and achieving the detection and characterization of the tensors from the early Universe at all the angular scales in which their signal is significant in CMB anisotropies. 

%
%
%
\bibliographystyle{spbasic}
\bibliography{opisiii}
\end{document}